\documentclass[%
 reprint,
 amsmath,amssymb,
 aps,
]{revtex4-1}

\usepackage{graphicx}
\usepackage{dcolumn}
\usepackage{bm}
\usepackage{xspace}
\usepackage{todonotes}

\begin{document}

\preprint{APS/123-QED}

\title{Proposal for a fully decentralized blockchain and proof-of-work algorithm for solving NP-complete problems}%
\thanks{All authors contributed equally to this work.}
\author{Carlos G. Oliver}
 \email{cgoliver@protonmail.com}
\affiliation{%
School of Computer Science, McGill University, Montreal, Canada
}%

\author{Alessandro Ricottone}
\author{Pericles Philippopoulos}
\affiliation{Department of Physics, McGill University, Montreal, Canada
}%

\date{\today }
\begin{abstract}

\end{abstract}

\maketitle


\emph{Abstract} We propose a proof-of-work algorithm that rewards blockchain miners for using computational resources to solve NP-complete puzzles. The resulting blockchain will publicly store and improve solutions to problems with real world applications while maintaining a secure and fully functional transaction ledger. 

\section{\label{sec:level1}Introduction}

The widespread success of cryptocurrency platforms such as Bitcoin  \cite{nakamoto2008bitcoin}, Ethereum \cite{wood2014ethereum} has attracted a substantial amount of computational resources ~\cite{o2014bitcoin}. However, the majority of this computing power is destined for executing proof-of-work algorithms (for example, the Bitcoin hashrate is over four exahash per second ~\cite{Redman}). While proof-of-work algorithms are highly reliable, the information generated by mining does not extend beyond guaranteeing the validity of the information on the network. In this work, we aim to design a new mining paradigm that diverts some of the computational resources from mining to solving problems with real world applications while simultaneously maintaining a secure blockchain.

We focus on the class of problems known as NP-complete problems. Such problems are most readily applied to blockchain systems because they possess the important property that the solution to the problem can be verified in polynomial time, while identifying the solution in the first place has no known polynomial algorithm. One of the first systems to attempt such a paradigm was Primecoin \cite{king2013primecoin}, which in 2013 proposed that users devote their computational power towards finding specific chains of prime numbers instead of cryptographic mining. While the identification of prime numbers is of interest generally, technical limitations force the coin to solve for a specific type of prime number whose scientific impact is not clear at the moment. Other blockchains such as the CureCoin~\cite{foldingcoin} (previously known as Foldcoin) and Coinami~\cite{ileri2016coinami} have attempted to solve bioinformatics problems which are of more direct impact. However, both systems depend on a central authority to delegate the problems and validate the identity of users who must register. We therefore aim at developing a cryptocurrency that addresses these shortcomings while still reliably acting as a trusted distributed ledger. The key components we aim to implement are:

\begin{itemize}
	\item Full decentralization 
	\item Anonymity
	\item Generation of solutions to practical problems
\end{itemize}

By proposing a novel mining and incentive protocol to the established Bitcoin framework, we believe we have achieved a system that satisfies all of the above features.

\section{Blockchain}

The proposed blockchain is based closely on the well established Bitcoin blockchain protocol, with one key difference: miners are rewarded for solving an NP-complete problem via a mining difficulty reduction. The idea is that miners, after building a block are allowed to choose between submitting a block mined using the standard Bitcoin protocol (see ~\cite{nakamoto2008bitcoin}) by finding a valid nonce, or if they have computed an improved solution to an NP-complete problem can publish it along with their block which would be verified and accepted by the network at a reduced difficulty.

For the sake of simplicity, we assume that the blockchain will deal with a single instance of a given NP-complete problem, we call this problem $P$. For example, if the NP-complete problem were finding a graph coloring (GC) of a graph $g$ of size less than some current best $k$, the blockchain would only work on that specific $g$ and $P := (g, k)$. In Section ~\ref{sec:problems} we discuss ways to incorporate a greater number of instances of the problem. The only requirement for a problem to be used in the proof-of-work is that verifying a solution to the problem can be accomplished in polynomial time with a clearly defined scoring scheme. 

\subsection{Block structure}

Blocks are structured identically to Bitcoin blocks (i.e. Transaction Merkle Root, miner address, nonce, etc.), with the addition of a new field for storing the problem, $P$. The genesis block is special in that it contains a field that stores the initial problem state. For example, if the blockchain is to solve the Traveling Salesman Problem (TSP), the block will simply store the graph to be worked on. Typically such problems can be compactly represented in matrix form. 

Subsequent blocks contain a field that stores a compact representation of a solution. In the case of TSP, this would be a vector with the index of visited nodes. However, not all blocks are forced to contribute a solution to the given problem, and this field is allowed to take on a null value that indicates that the miner produced the block normally.

\subsection{Incentivization}

The main task a Bitcoin miner has to accomplish in order to produce a valid block once he has obtained a set of valid transactions is to find a valid nonce, $n$. More specifically, miners need to find an integer $n$ such that 
\begin{equation}
H(B, n) < \epsilon_d,
\end{equation}
where $H(B,n)$ is the hash function applied to block $B$ and $n$, and $\epsilon_d$ is the target associated with difficulty $d$.  Because $H$ is a cryptographic hash function, the output of $H$ is completely unpredictable, therefore mining consists of brute forcing values of $n$ until the condition is satisfied. Clearly an $\epsilon_d$ with smaller $d$ is easier to satisfy than the inverse. Once a miner finds a valid nonce they can publish the block along with the $n$ and the network can easily verify the validity of the block according to the network's current difficulty $d$. Because mining is a competitive endeavor, we reward miners that compute solutions to $P$ by reducing mining difficulty. In our protocol, miners will accept blocks that satisfy a reduced difficulty $d_r$ constraint if the block contains a solution that is better than the current best solution on the blockchain. If the average time needed to mine a block \emph{with} a solution is shorter than the average time needed to mine a block \emph{without}, we expect miners to spend their computational power in improving the solution to the problem.
 
\subsection{Mining \& Difficulty Scaling}

In Bitcoin, the hashing difficulty is retargetted every $N$ $(=2016)$ blocks, by comparing the time it took to mine these blocks, $T^{\star}$, to a target time $T$. In the proposed blockchain, we have two different difficulties which must each be retargetted. These difficulties are $d_b$, for blocks mined \emph{without} a solution to problem $P$, and $d_r$, a reduced difficulty for blocks mined \emph{with} a solution to problem $P$. For the retargetting of both difficulties to be accomplished, a value for $T$ must be fixed by the blockchain, as is the case for Bitcoin. However, we must also establish the constant $0<\eta<1$, which represents the desired ratio of time it takes to mine a block with a solution to problem $P$, $t_s$, to the time it takes to mine a block without a solution to problem $P$, $t_b$. A smaller $\eta$ would result in a greater incentive for puzzle solutions to be found.

We define $t_s^{\star}$ to be the average time it took to mine a block with a solution to problem $P$ and $t_b^{\star}$ to be the average time it took to mine a block without a solution to problem $P$. We can thus write these two quantities as

\begin{eqnarray} 
&t_s^{\star} = \frac{d_r}{p_H} + \frac{d_p}{p_H}, \label{eq:ts} \\
&t_b^{\star} = \frac{d_b}{p_H}, \label{eq:tb}
\end{eqnarray}
where $p_H$ is the rate at which computations can be performed by the network, and $d_p$ is the difficulty associated with solving the problem $P$. We expect $d_p$ to increase with time as better and better solutions to problem $P$ are found. We also define the \emph{measured} quantity, $\eta^{\star} = t_s^{\star}/t_b^{\star}$ and $b$, the fraction of blocks mined with base difficulty $d_b$. Using Eqs. (\ref{eq:ts}) and (\ref{eq:tb}), 
\begin{equation}
\eta^{\star} = \frac{d_r + d_p}{d_b}.
\end{equation}

The goal, after $N$ blocks, is to set $d_r$ and $d_b$ to new values $d_r'$ and $d_b'$ so that the time it takes to mine the $N$ blocks, $T^{\star}$, under fixed $p_H$, $d_p$ and $b$, readjusts towards the target $T$ and $\eta^{\star}$ readjusts towards $\eta$. We can therefore write
\begin{eqnarray}
T^{\star} = \frac{N}{p_H}d_b \left[b + (1-b)\eta^{\star}\right], \\
T = \frac{N}{p_H}d_b' \left[b + (1-b)\eta\right].
\end{eqnarray}
Solving these equations for the retargetted base difficulty yields,
\begin{equation}\label{eq:dbretarget}
d_b' = d_b \left[\frac{b + (1 - b) \eta^{\star}}{b + (1 - b) \eta} \right]\frac{T}{T^{\star}}.
\end{equation}

To retarget $d_r$ we use the fact that the difficulties should be updated so that
\begin{equation}\label{eq:eta}
\eta =  \frac{d_r' + d_p}{d_b'}.
\end{equation}

We note that $d_p$  is not retargetted to $d_p'$ as it is completely determined by the status of the problem $P$. Combining Eq. (\ref{eq:eta}) with Eqs. (\ref{eq:ts}) and (\ref{eq:tb}), we can write 
\begin{equation}\label{eq:drretarget}
d_r' - d_r = \eta d_b' - \eta^{\star} d_b.
\end{equation}
Thus, Eqs. (\ref{eq:dbretarget}) and (\ref{eq:drretarget}) display the rules for the difficulty retargetting of the proposed blockchain. As with the Bitcoin blockchain, it would be advisable to introduce a maximum retargetting factor (4 for the Bitcoin chain) to avoid a change in difficulty that is too abrupt. In other words, we would enforce

\begin{equation}
\frac{1}{4}\le \frac{d_{b}'}{d_{b}}\le 4.
\end{equation}

After $P$ has become too difficult to improve, it is possible that $d_p > d_b$. This implies that the updated difficulty $d_r'$ becomes negative at which point it would be necessary to introduce a new problem $P$ to work on. We discuss how this can be implemented in  Section ~\ref{sec:problems}. However, at the point when it is no longer worthwhile to solve the puzzle, the blockchain naturally functions with $d_b$ and the standard Bitcoin proof-of-work.
  
\subsection{Possible Puzzles}

There is a wealth of interesting NP-complete problems that could be used as puzzles for the blockchain. Here we suggest a list, which is by no means exhaustive, of possible applications:

\begin{itemize}
	\item Multiple Sequence Alignment: many databases with DNA sequence information are widely available (e.g. ~\cite{sherry2001dbsnp}). The problem of improving aligned multiple DNA sequences is NP-complete and has many applications in biology and medicine.
    \item Protein/biomolecule folding and design: computing the 2D and 3D geometry of chains of DNA/RNA/Protein as well as designing sequences with desired geometries. Many databases for this problem are also available, e.g. Protein Data Bank ~\cite{berman2006protein}. Solutions so this problem can have direct medical applications.
    \item Ising-lattice: The decision form of the Ising model (to decide whether the ground state of an Ising Hamiltonian has energy $E\le 0$) is NP-complete ~\cite{lucas2013ising} and it can be mapped to many other NP-complete problems. These models are widely studied in physics.
\end{itemize}

\section{Multiple-Puzzle Blockchain}\label{sec:problems}

\begin{figure*}
	\includegraphics[width=0.9\textwidth]{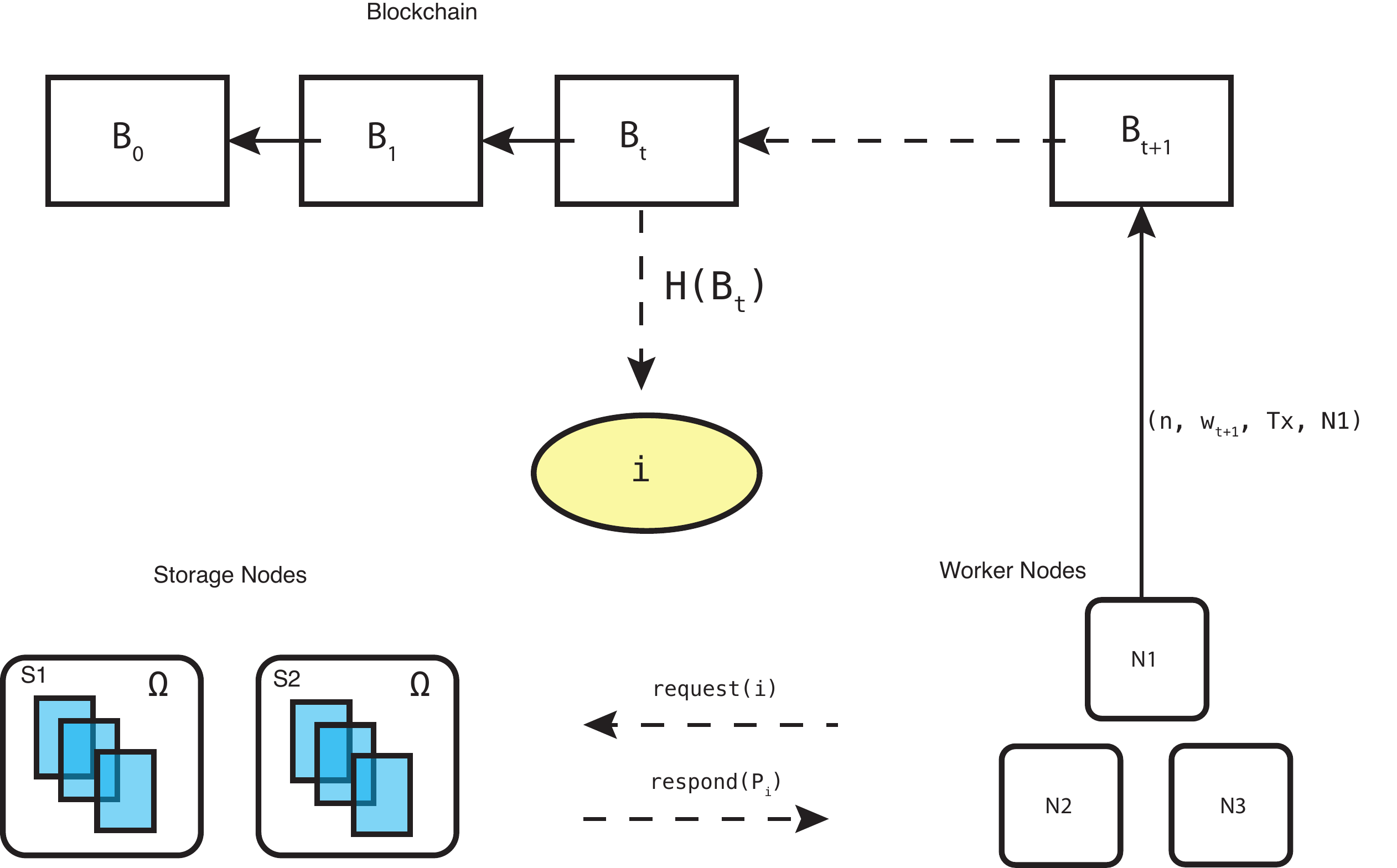}
	\caption{Potential blockchain architecture for handling multiple puzzles. Problem set $\Omega$ is stored in independent storage nodes with addresses S1 and S2 (lower left) that sync to the main blockchain (top). Worker nodes N1, N2, N3, that mine coins and solve puzzles request puzzle data from the storage nodes. The index(es) $i$ of the admissible puzzle(s) for block $B_{t+1}$ are obtained from the hash of the block $H(B_t)$, which is also contained in $B_{t+1}$. The hash-determined problem index is an optional feature and can be omitted if the blockchain allows any problem solution to be accepted at any time. Once a miner has successfully mined a block with transactions \texttt{Tx} by finding a valid nonce \texttt{n} and puzzle solution $w_{t+1}$ he publishes the block to the network.}
    \label{fig:diagram}
\end{figure*}

Clearly a blockchain that only works on a single problem, $P$ will quickly exhaust its usefulness. Ideally, we would want the blockchain to simultaneously solve a set $\Omega \subseteq \Omega_j$ of instances of an NP-complete problem $j$, where $i$ is an index over all possible NP-complete problems, $j \in \{ \mathrm{GC}, \mathrm{TSP}, \dots\}$, and $\Omega_j$ contains all possible NP-complete problems of type $j$. While we leave the specifics of how this can be achieved to future work, here we discuss some of the challenges and potential solutions. 

\subsection{Puzzle Storage}

Because we wish to have full decentralization, $\Omega$ must be stored on the blockchain. This can be achieved in one of two manners. If the size of $\Omega$ is small enough to be stored by all the miners, the genesis block could simply be used to store an indexed database of $\Omega$. Subsequent blocks that contain solutions to a problem $P \in \Omega$ would simply include with their solution a pointer to the corresponding problem in the genesis block. If $\Omega$ is too large for all the miners to store, the network could instead allow some nodes to participate as `storage' nodes and collect a reward for doing so in a manner similar to file storage coins ~\cite{FileCoin}. The main blockchain would then simply contain pointers to the relevant problem for each block ({\bf Fig. ~\ref{fig:diagram}, lower half)}.

\subsection{Puzzle Selection}

In the single-puzzle blockchain, miners always work on the same problem. In the multiple-puzzle setting there must be a protocol for selecting the problem for the current block. The most natural approach is to allow miners to submit a solution to any $P \in \Omega$ they choose for a reduced difficulty. This has the advantage of potentially maximizing the network's efficiency in solving problems if miners work on non-overlapping regions of $\Omega$. However, if one wishes to force some distribution on the frequency that each puzzle gets included in a block, one could use the hash of the previous block to determine the index $j$ of the current admissible puzzle, $P_i \in \Omega$. ({\bf Fig. ~\ref{fig:diagram}, upper half}) Such a function could be augmented for example to favor problems not yet been included and suppress problems that have been worked on too much. 

\subsection{New Puzzles}

Eventually (unless $\Omega$ is very large), all puzzles will reach some plateau of optimality and further computation will not produce significant gains. The obvious solution to this situation is to allow the blockchain to incorporate new problems into $\Omega$. The challenge is to manage new puzzle incorporation without a central authority that ensures problems are: 1) valid instances of $j$ 2) of interest 3) not already solved. New puzzles can be included as a special transaction. This transaction can store the problem in the current block, or add it to the storage nodes. To discourage puzzles that do not satisfy the three criteria, new problems should be submitted with a fee which is then distributed as a reward to miners that solve the puzzle. Alternatively, the network can agree on a new problem set off-chain and in a manner similar to Bitcoin, induce a fork or upgrade of the chain that includes a new agreed upon problem set.

\section{Conclusion}

We have outlined the major components of a cryptocurrency system that incentivizes the identification of solutions to scientifically interesting problems without relying on any central authorities or servers. The benefit of using this proposed blockchain instead of other proof-of-work blockchains is twofold. A portion of the power expended on hashing is redirected to solving problems that are scientifically relevant. And second, solutions to these problems are naturally stored and updated in the blockchain for public access.  To our knowledge, this is the first blockchain solution that achieves these features in a fully decentralized manner. The implementation of this system is left for future work.

\bibliography{biblio}
\end{document}